\documentclass{elsart}
\usepackage{psfig,natbib,epsfig,graphicx}
\usepackage{amssymb}
\journal{Nuclear Physics}
\begin{document}

\begin{frontmatter}

\title{On the Excitation of Double Giant Resonances in Heavy Ion Reactions}

\author[sevilla]{C.H. Dasso},
\author[padova]{L. Fortunato},
\author[catania]{ E. G. Lanza} and
\author[padova]{A. Vitturi}

\address[sevilla]{Departamento de F\'{\i}sica At\'omica, Molecular y Nuclear,\\ Universidad
de Sevilla, Spain}
\address[padova]{Dipartimento di Fisica ``G.Galilei", Universit\'a di Padova and INFN,
 Italy}
\address[catania]{INFN-Catania and Dipartimento di Fisica e Astronomia,\\  Universit\'a di Catania, Italy}

\begin{abstract}
The interplay of nuclear and Coulomb processes in the inelastic excitation
of single- and double-phonon giant resonances in heavy ion collisions
is studied within a simple reaction model.
Predominance of the Coulomb excitation mechanism on the population of the 
single-phonon and, on the contrary, predominance of the nuclear excitation
for the double-phonon is evidenced.  
The effect of the spreading of the strength distribution of the giant 
resonances on the excitation process is analyzed, showing sizeable 
modifications in the case of Coulomb dominated processes.

\end{abstract}
\begin{keyword}
Multiphonon collective excitation, Double Giant Resonance, Spreading
width.
\PACS  24.30.Cz, 25.70.-z, 25.70.De
\end{keyword}
\end{frontmatter}

\section{Introduction.}
Giant Resonances (GR) are considered as one of the most important
elementary modes of excitation in nuclei~\cite{harak}. They are
generally interpreted as harmonic density vibrations around the
equilibrium distribution of the nucleons. Within this point of view
one should also expect to observe higher-lying states of the harmonic
spectrum such as, for instance, the two-phonon Double Giant Resonance
(DGR). The existence of the double-phonon excitation of low-lying
collective vibrational states has been known for a long time, but it
is only recently that the multiple excitations of GR have been
systematically observed (for a complete review see ref.\cite{rev} and
\cite{harak} and references therein).

Interest in this subject has been renewed by recent experiments with
relativistic heavy-ion beams, where inelastic cross sections for the
excitation of the dipole DGR have been precisely measured. The
theoretically calculated cross sections --~when performed within the
framework of the standard harmonic model~-- systematically
underestimate the experimental data by as much as a factor of
two. This unexpected enhancement of the cross sections puts in
evidence shortcomings in either the description of the structure of
the modes or in the formulation of the reaction mechanism. Attempts to
improve over this situation have followed different paths.

The microscopic understanding of these resonances, for instance, has
been taken beyond the simple superposition of the 1p-1h configurations
to include couplings to 2p-2h, 3p-3h and/or states of higher
complexity~\cite{wam,pon,ber}. Residual interactions give rise to
anharmonicity in the energy spectrum~\cite{cat} and, also, appreciable
changes in the structure of the wave functions.  Recently, a
systematic study of the anharmonicity in the dipole DGR has been
carried out for several nuclei~\cite{pon2}. This study, based on a
quasiparticle RPA, has shown an effect of few hundred keV. The
same order of magnitude had been found in ref.~\cite{lan} for
\nuc{208}{Pb} and \nuc{40}{Ca}.  These effects have been taken into
account in macroscopic models that add small anharmonic
contributions~\cite{vol,bor} to the otherwise harmonic hamiltonian in
the presence of an external time-dependent field. Depending on the
magnitude of these anharmonic terms the inelastic cross sections for
the population of the dipole DGR can reach values which are close to
the experimental data. Microscopic calculations in the context of the
RPA approximation, have also succeeded in reducing the discrepancy
between the experimental data and the theoretical predictions down to
the level of a few per cent~\cite{lan}. Another approach to the
problem that has been examined~\cite{car,wei} exploits the so-called
Brink-Axel hypothesis~\cite{brax}. It also seems possible, through
this formalism, to obtain enhancements in the population of states in
the energy range around the DGR.

In this paper we set to investigate the role of the nuclear coupling
in the excitation of GR's and DGR's and its interplay with the
long-range Coulomb excitation mechanism.  Furthermore, we study the
consequences of the spreading of the strength distribution of the
single giant resonance on the inelastic cross section for both the GR
and DGR. These topics have been previously explored in the literature.
In refs.~\cite{cat87,lan2,and} nuclear and Coulomb interactions where
taken into account for medium-heavy nuclei at low bombarding energy
(around 50 MeV/A).  While these studies put in evidence interference
effects between the two excitation mechanisms there was no clear
resolution concerning what could be actually attributed to each of
them.  Also, the role played by the resonances' width on the reaction
cross sections was covered in refs.~\cite{can,ber96,wei}. The
analyses, however, were done only for the case of the Coulomb
excitation mechanism and lead to somewhat ambiguous results.

We shall do this survey within a simple reaction model that has the
virtue of conforming to the standard treatment of inelastic
excitations which is familiar to many active participants in this
field.  Our original intention was limited to investigate the
qualitative dependence of the probabilities of excitation of the
Double Giant Resonances as a function of several global parameters
such as the excitation energies, bombarding energies, multipolarity,
anharmonicity, width, etc.  In the process of refining the computer
programs we used to obtain these global trends we ended up with a
quite transparent and yet powerful tool that --~we believe~-- can be
useful for the experimentalists to make {\it quantitative} predictions
for measurements in a wide variety of circumstances.  With this very
practical purpose in mind we shall take in this contribution the width
of the states as a free parameter.  We shall also limit our
calculations to the non-relativistic regime and, for the different
examples, consider the excitation of single- and double-phonon Giant
Quadrupole Resonances.
 
Following this Introduction we describe in Sect. II the formalism
employed to make our estimates.  Relevant results for the reaction
$^{40}$Ar + $^{208}$Pb are given with an abundance of illustrations in
Sect. III.  The conclusions that can be inferred from these examples
are also the subject of this Section. Some concluding remarks are left
for Sect. IV.

\section{The Model}

The excitation processes of the one and two-phonon states are
calculated within the framework of the standard semiclassical model of
Alder and Winther \cite{ald} for energies below the relativistic
limit. According to this model for heavy ion collisions, the nuclei
move along a classical trajectory determined by the Coulomb plus
nuclear interaction. We will explore the energy range from few MeV up
to hundreds of MeV per nucleon. During their classical motion the
nuclei are excited as a consequence of the action of the mean field of one nucleus 
on the other. The excitation processes are described according to quantum
mechanics and they are calculated within perturbation theory.

We assume that the colliding nuclei have no structure except for the
presence, in the target, of one and two-phonon states whose energies
are $E_1$ and $E_2=2 E_1$, respectively. For the ion-ion potential we
have used the Coulomb potential for point charged particles and the
Saxon-Woods parametrization of the proximity potential $U_N(r)$ that
are commonly used in heavy ion collisions~\cite{win}.

In the theory of multiple excitations the set of coupled equations
describing the evolution of the amplitudes in the different channels
can be solved within the perturbation theory. We can write the
probability amplitude to excite the $\mu$ component of the one-phonon
state with multipolarity $\lambda$ as

\begin{equation}
a^{(1)}_{\lambda \mu} (t)= (-i / \hbar) 
\int_{-\infty}^{\infty} dt F_{\lambda \mu }  
(r(t),\hat{r}(t))  e^{iE_1 t/\hbar}~,
\label{a1}
\end{equation}
where the integrals are evaluated along the classical trajectories
${\bf{r}}(t)$. In this equation the main ingredient is the coupling
form factor
\begin{equation}
 F_{\lambda \mu }  (r(t),\hat{r}(t))~=~ f_{\lambda}(r)~Y_{\lambda \mu}
(\hat{r})~,
\end{equation}
chosen according to the standard collective model
prescription~\cite{land}. For a given multipolarity $\lambda$ the radial
part assumes the form
\begin{equation}
f_\lambda (r) = {{3 Z_p Z_t e^2}\over {(2 \lambda +1) R_C}}
\beta^C_\lambda \Bigg( {R_C \over r} \Bigg)^{\lambda +1}
- \beta^N_\lambda R_T {d\over dr} U_N(r)~.
\label{ff}
\end{equation}
The deformation parameters $\beta$ determine the strength of the
couplings, and they are normally directly associated with the
$B(E\lambda)$ transition probability.  The expression for the nuclear
component of the form factor is not valid for $\lambda=1$. In this
case the inelastic form factor is obtained from the Goldhaber-Teller
or Jensen-Steinwendel models. The $Z_p$ ($Z_t$) denotes the charge
number of the projectile (target), while $R_C$ and $R_T$ are the
Coulomb and matter radii of the target nucleus.

In a similar way, the amplitude for populating the two-phonon state
with angular momentum $L$ and projection $M$ can be obtained as
\begin{eqnarray}
a^{(2)}_{LM} (t) &=& (1/ \hbar )^2
\sum_{\mu}~ \sqrt{(1 +\delta_{\mu,M-\mu})} \nonumber \\
& \times & \int_{-\infty}^{\infty} dt F_{\lambda,M-\mu}({\bf r}(t))  e^{i(E_2-E_1)t /\hbar}
\int_{-\infty}^{t} dt' F_{\lambda,\mu}({\bf r}(t'))  e^{iE_1t' /\hbar}
\label{a2}
\end{eqnarray}

Solving the classical equation of motion we can calculate for each
impact parameter $b$ the excitation probability $P^{(1)}(b)$ and
$P^{(2)}(b)$ to populate the single- and the double-phonon
state. These are given by
\begin{equation}
 P^{(1)}(b)~=~\sum_{\mu} |a^{(1)}_{\lambda \mu}(t=+\infty)|^2
\end{equation}
and
\begin{equation}
 P^{(2)}(b)~=~\sum_{L}P^{(2)}_{L}~=~\sum_{LM} |a^{(2)}_{LM}(t=+\infty)|^2~.
\end{equation}
In order to get the corresponding cross sections we have then to
integrate the probabilities $P^{(\alpha)}$'s ($\alpha$ =1,2)
\begin{equation}
\sigma_{\alpha} = 2 \pi \int_0^\infty P^{(\alpha)}(b) T(b) b db \,.
\label{xsec}
\end{equation}
Generally, in Coulomb excitation processes the transmission
coefficient is taken equal to a sharp cutoff function
$\theta(b-b_{min})$ and the parameter $b_{min}$ is chosen in such a
way that the nuclear contribution is negligible. We want to take into
account also the contribution of the nuclear field so in our case T(b)
should be zero for the values of $b$ corresponding to inner trajectory
and then smoothly going to one in the nuclear surface region. This can
be naturally implemented by introducing an imaginary term in the
optical potential which describes the absorption due to non elastic
channels.  Then the survival probability associated with the imaginary
potential can be written as
\begin{equation}
T(b)= \exp\Bigg\{ {2\over \hbar }  \int_{-\infty}^{+\infty} W(r(t)) dt \Bigg\} \,,
\label{Tb}
\end{equation}
where the integration is done along the classical trajectory. The
imaginary part $W(r)$ of the optical potential was chosen to have the
same geometry of the real part with half its depth.

The excitation processes of both single and double GR can change
significantly when one takes into account the fact that the strength
of the GR is distributed over an energy range of several MeV. Among
the few standard choices for the single GR strength distribution, we will
assume a Gaussian
shape, with a width $\Gamma_1=2.3 \sigma $ which we will take as a
parameter, of the following form
\begin{equation}
S(E) = {1 \over {\sqrt{2 \pi}\sigma}} \exp \Bigg\{ 
{-(E - E_1)^2 \over {2 \sigma^2}}\Bigg\} \,.
\label{GAU}
\end{equation}
Calculations have been also performed with a Breit-Wigner shape
yielding similar trends. However, the Gaussian form guarantees a
better localization of the response and prevents superposition of the
modes for the largest widths (for a further discussion see
ref.~\cite{ber}). 

To get the cross section to the one-phonon state one then defines a
probability of excitation per unit of energy,
\begin{equation}
 dP^{(1)}(E,b)/dE~=S(E)~\sum_{\mu} |a^{(1)}_{\lambda \mu}(E,t=+\infty)|^2~,
\end{equation}
where the single amplitudes $a^{(1)}_{\lambda \mu}(E,t)$ are obtained
as before, but with a variable energy $E$.  The probability of
exciting the double-phonon state is then obtained by folding the
probabilities of single excitation, in the form
\begin{equation}
dP^{(2)}(E,b)/dE~=~\int dE'~ {dP^{(1)}(E',b)\over dE}~ ~ 
{dP^{(1)}(E-E',b)\over dE}~. 
\end{equation}
The total cross section for one- and two-phonon states can then be
constructed as
\begin{equation}
\sigma_{\alpha} = 2 \pi \int_0^\infty \int_0^
\infty {dP^{\alpha} \over dE}(E,b)~ T(b) \,b \,db
\,dE \, \,.
\label{xsecg}
\end{equation}

Due to the Q-value effect it is clear that one expects a distortion in
the shape of the distribution of the cross section which will favor
the lower part of the distribution in energy.

\section{Results}

We show in Fig.~1 the dependence on the impact parameter of the
excitation probabilities for the one- and two-phonon states of the
Giant Quadrupole Resonance in lead.  The reaction we have chosen for
this illustration is $^{40}$Ar + $^{208}$Pb at a bombarding energy of
40 MeV per nucleon.  The deformation parameters have been chosen equal
$\beta^C = \beta^N = 0.07$, in agreement with the currently estimated
value for the $B(E2)$. The range of impact parameters given in the
figure covers the relevant grazing interval, and in a classical
picture (including both Coulomb and nuclear fields) 
yields scattering angles between $3.4$ and $5.5$ degrees.  In
the strictly harmonic case the probabilities for excitation of the
double-phonon state can of course be constructed from those
corresponding to the single-phonon; they are both explicitly given
here for a matter of later convenience.  Each frame displays a set of
three curves that allows us to compare the individual contributions of
the Coulomb and nuclear fields to the excitation process and put in
evidence a value of $b\approx 12.5$ fm for the maximum (destructive)
interference between the competing mechanisms.

\begin{figure}[!t]
\begin{center}
\epsfig{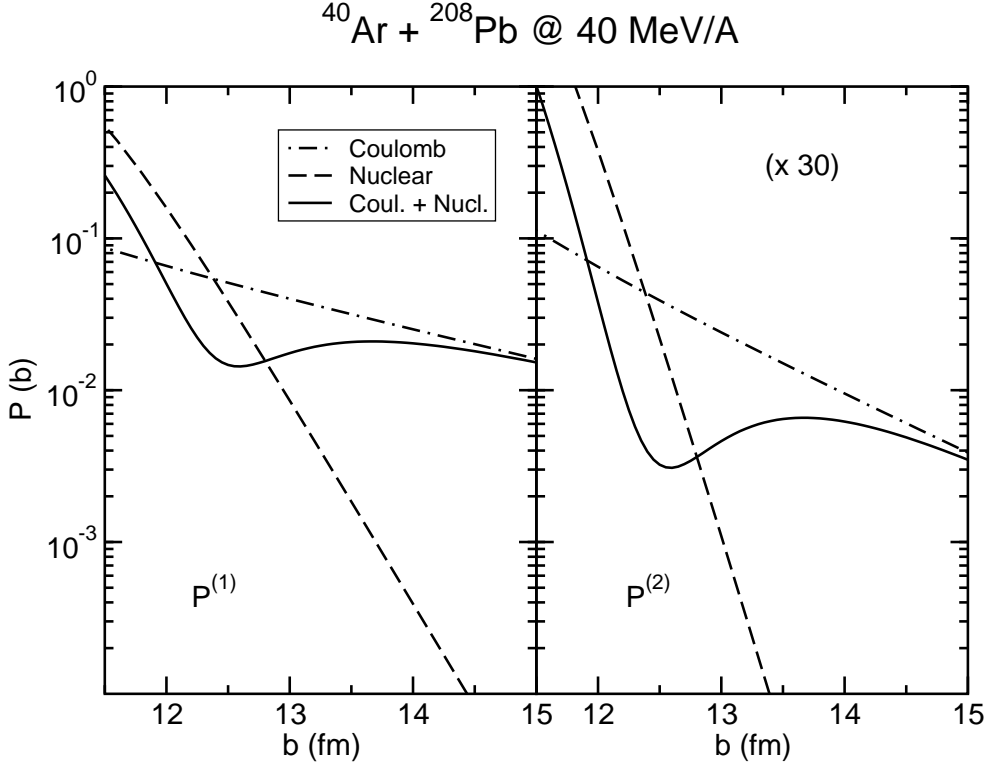}
\end{center}
\caption{Excitation probability vs. impact parameter for the one- (left part)
and two-phonon (right part) states of the GQR in lead for the reaction
$^{40}$Ar + $^{208}$Pb at 40 MeV/A. The Coulomb (dot-dashed line) and
nuclear (dashed) probabilities are displayed as well as the total
(solid line). The curves on the right part have been multiplied by
30.}
\label{PB}
\end{figure}

\begin{figure}[!t]
\begin{center}
\epsfig{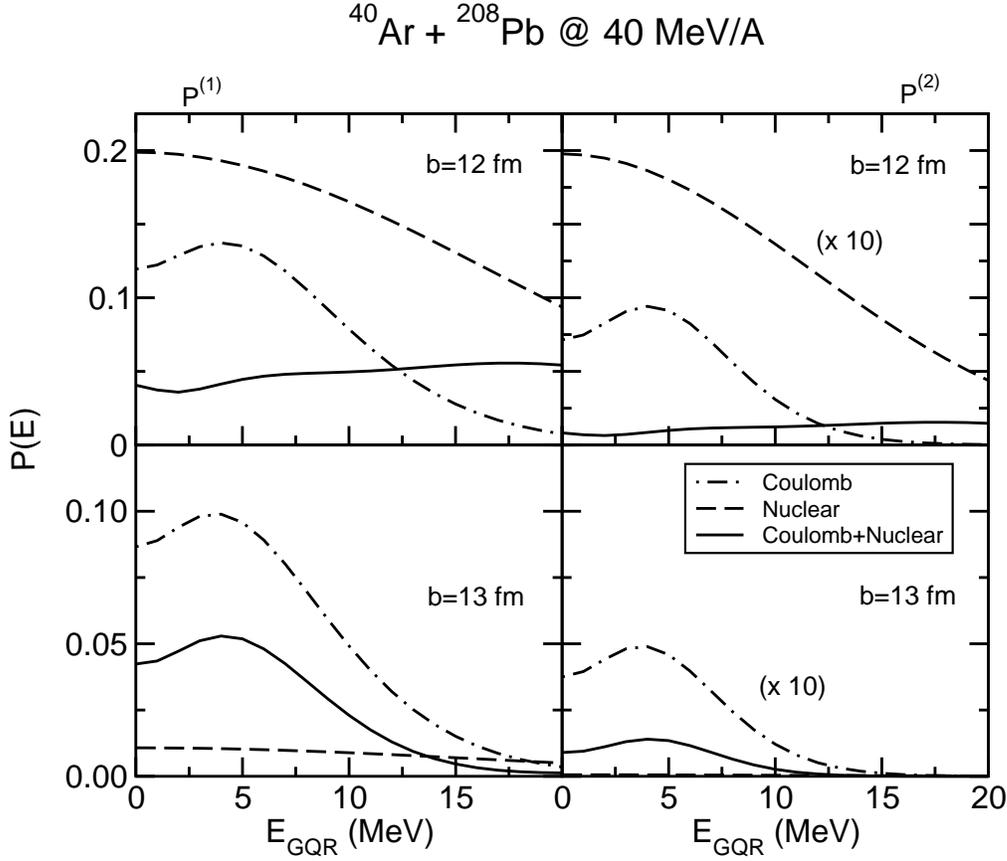}
\end{center}
\caption{ Excitation probabilities as function of the GQR energy, which
is taken as a parameter, for the reaction $^{40}$Ar + $^{208}$Pb at 40
MeV/A. The graphs on the left correspond to the excitation probability
of the single GQR ($P^{(1)}$) while the ones on the right correspond
to DGQR ($P^{(2)}$) and they are multiplied by a factor 10. The
Coulomb (dash-dotted line) and nuclear (dashed) probabilities are
displayed as well as the total (solid line). The upper (lower) figures
correspond to an impact parameter of 12 (13) fm.}
\label{PQ}
\end{figure}

\begin{figure}[!t]
\begin{center}
\epsfig{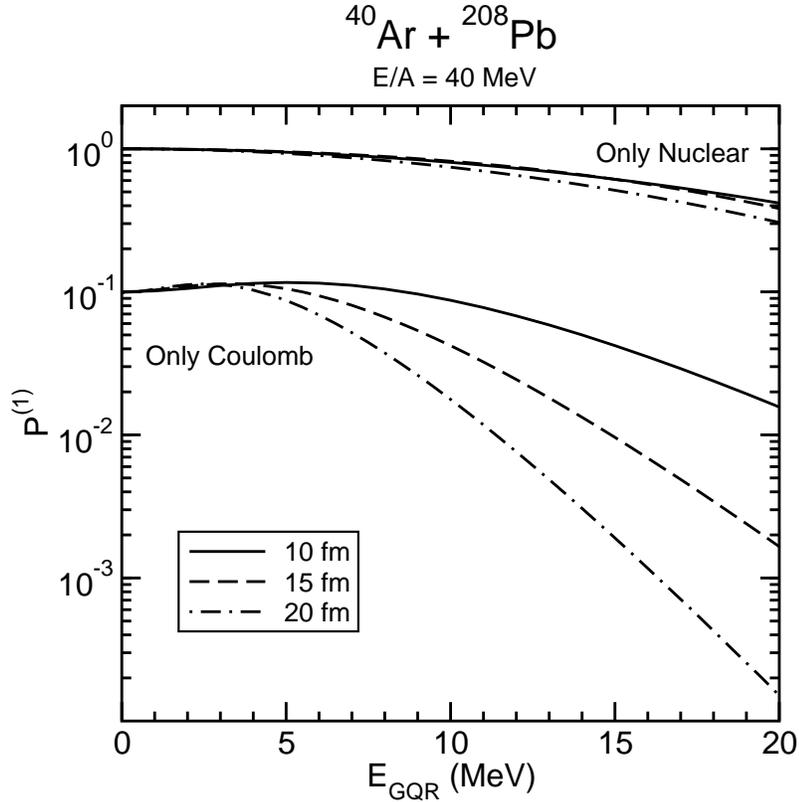}
\end{center}
\caption{ Excitation probability for the single GQR as a function of the
GQR energy for three values of impact parameters. They have been
normalized to their values at $E_{GQR} =0$.  The upper curves
correspond to the excitation probability due only to the nuclear
field. The probabilities calculated only with the Coulomb field are
shown in the lower part of the picture. They have been divided by 10
in order to render the figure readable.}
\label{P1QB}
\end{figure}

\begin{figure}[!t]
\begin{center}
\epsfig{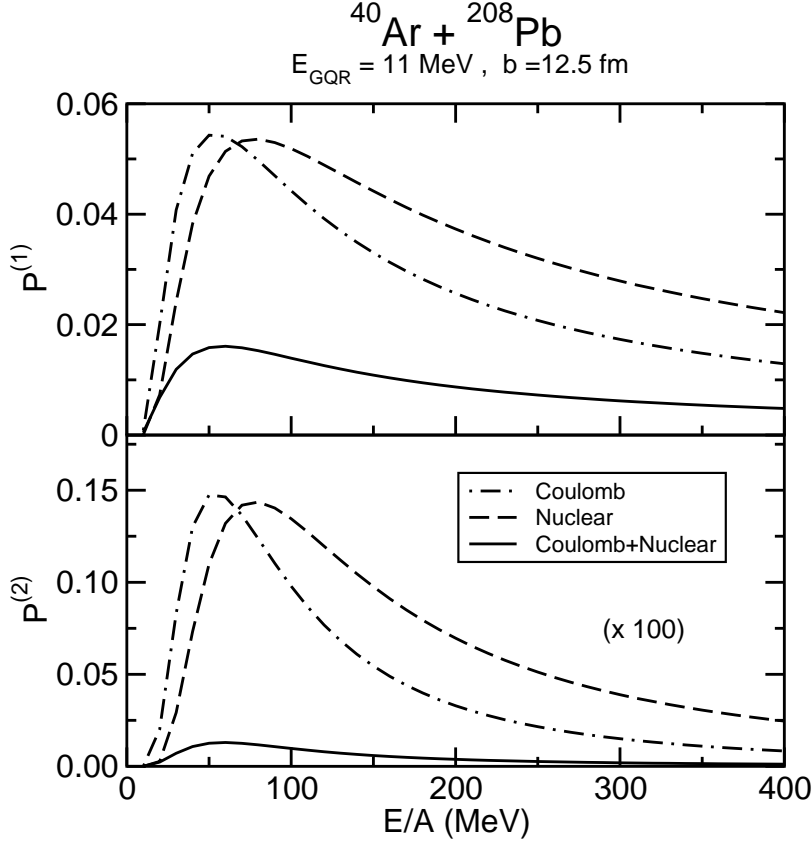}
\end{center}
\caption{Excitation probability as a function of incident energy of the one- (upper part)
and two-phonon (lower part) states of the GQR in lead for the reaction
$^{40}$Ar + $^{208}$Pb for an impact parameter of 12.5 fm. The Coulomb
(dot-dashed line) and nuclear (dashed) probabilities are displayed as
well as the total one (solid line). The curves on the lower part have
been multiplied by 100.}
\label{PE}
\end{figure}

\begin{figure}[!t]
\begin{center}
\epsfig{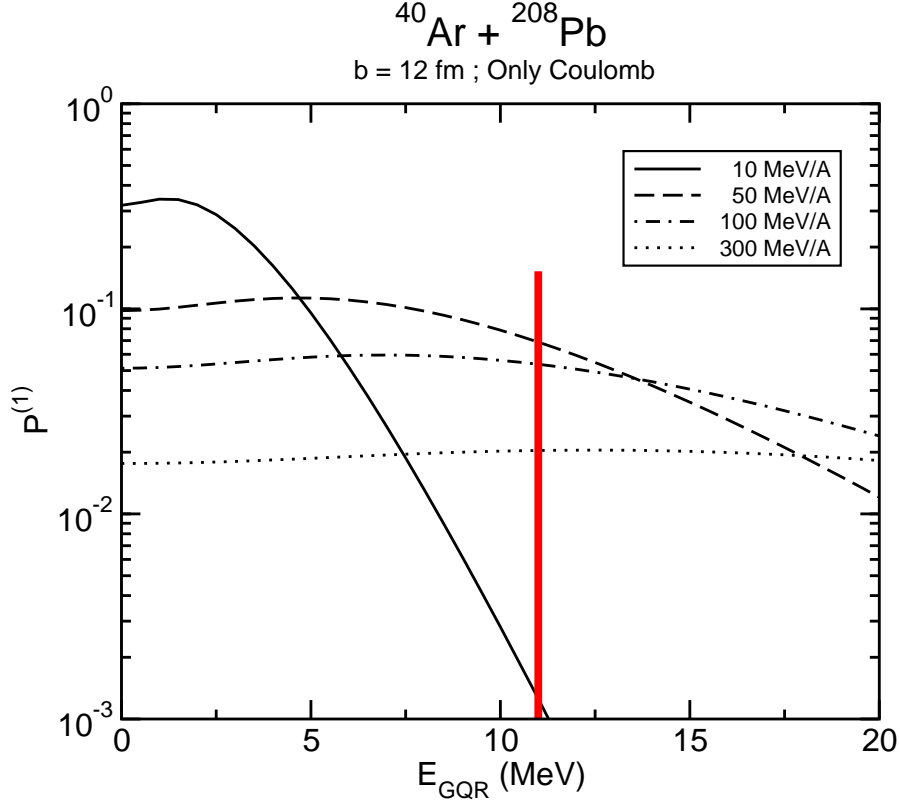}
\end{center}
\caption{Coulomb excitation probability for the one-phonon state 
as a function of the Q-value and for four different incident energies
as shown in the legend. The curves correspond to calculations done for
an impact parameter b=12 fm. The vertical line indicates the GQR
energy for lead.}
\label{P1QE}
\end{figure}

\begin{figure}[!t]
\begin{center}
\epsfig{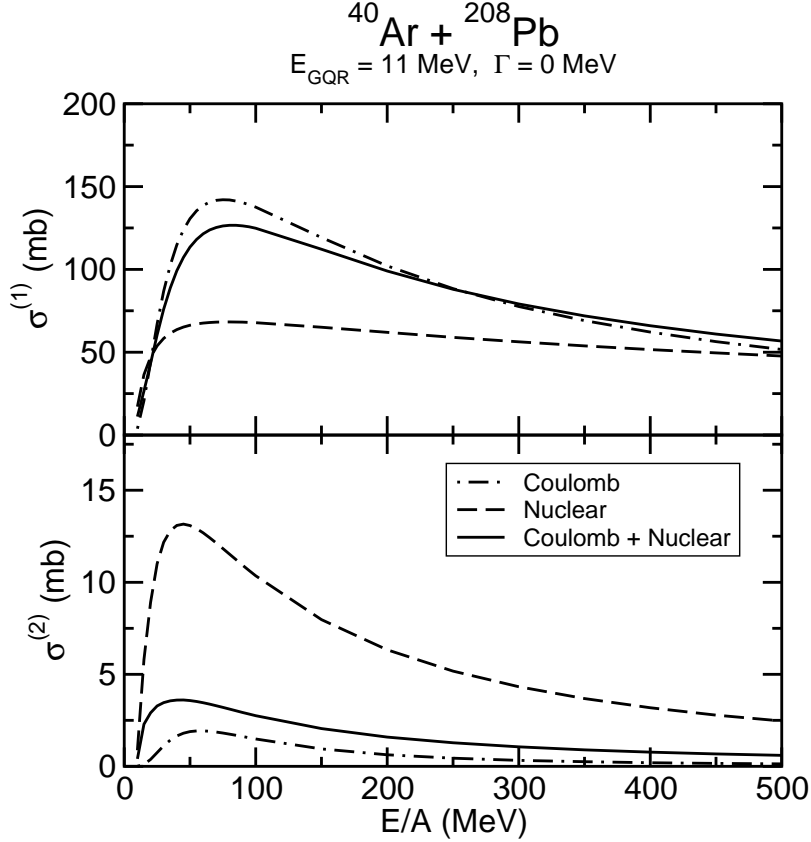}
\end{center}
\caption{Excitation cross section for the GQR (upper figure) and DGQR
state (lower figure) as a function of the incident energy. Again, the
Coulomb (dot-dashed line), nuclear (dashed line) and total (solid
line) contributions are shown.}
\label{XE}
\end{figure}

\begin{figure}[!t]
\begin{center}
\epsfig{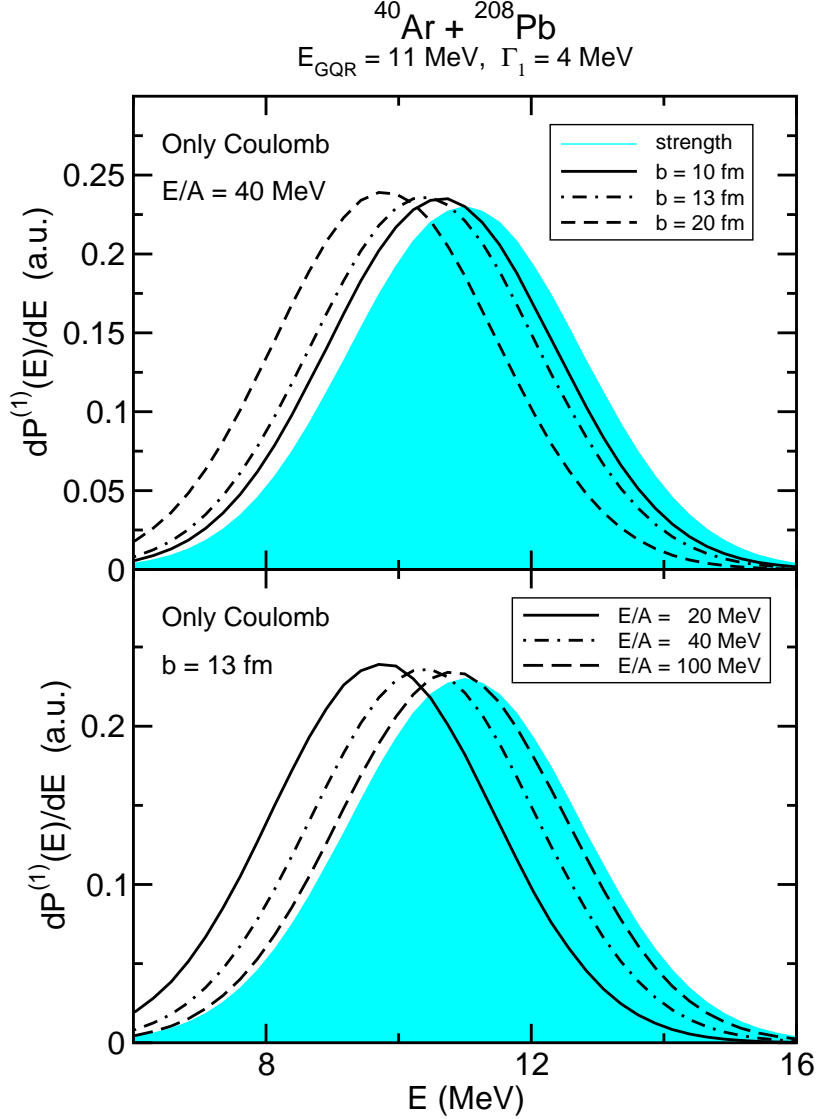}
\end{center}
\caption{Normalized distributions of Coulomb excitation probability for the one-phonon state 
for different impact parameters (upper figure) and bombarding energies
(lower figure) as shown in the legend. The shaded area shows the
Gaussian strength distribution used as input in the calculation. The
width has been chosen to be 4 MeV.}
\label{P1W}
\end{figure}

\begin{figure}[!t]
\begin{center}
\epsfig{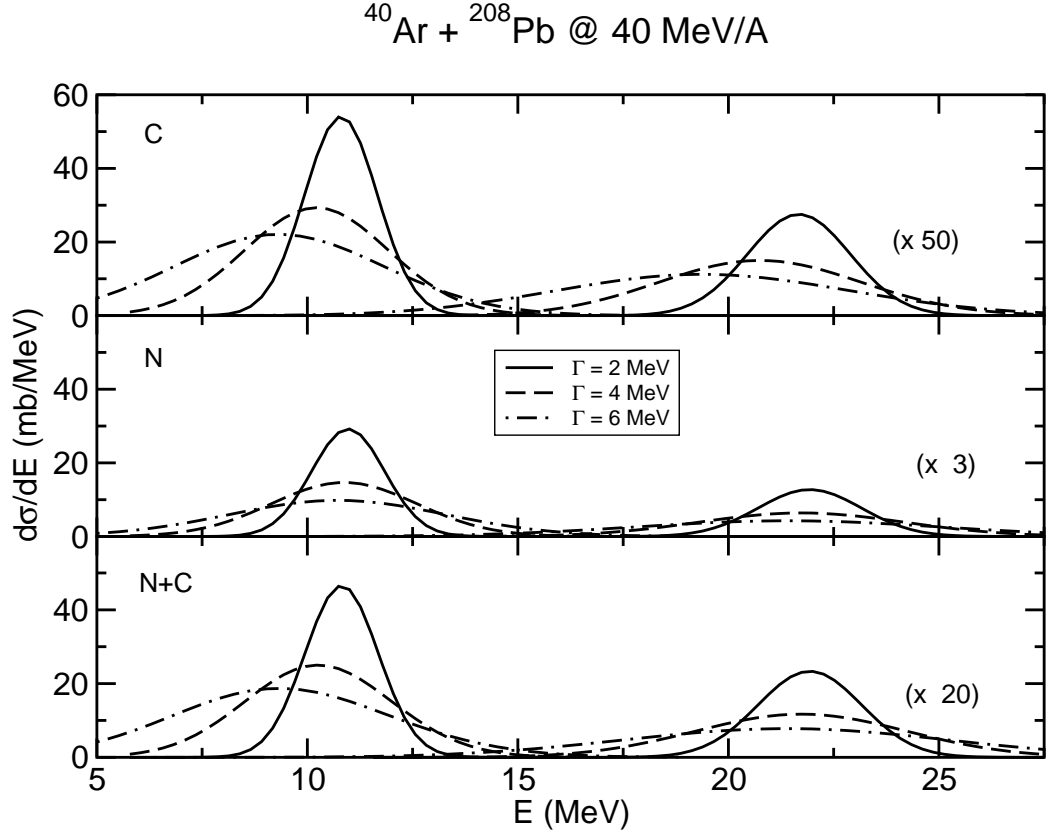}
\end{center}
\caption{Cross section distribution for the reaction
$^{40}$Ar + $^{208}$Pb at 40 MeV/A for three values of the width as
indicated in the legend.
The contributions of the Coulomb (C), nuclear (N) and total (N+C) cross
sections are shown in single graphs. The cross sections for DGQR are
multiplied by the factors reported in the figure.}
\label{XW}
\end{figure}

\begin{figure}[!t]
\begin{center}
\epsfig{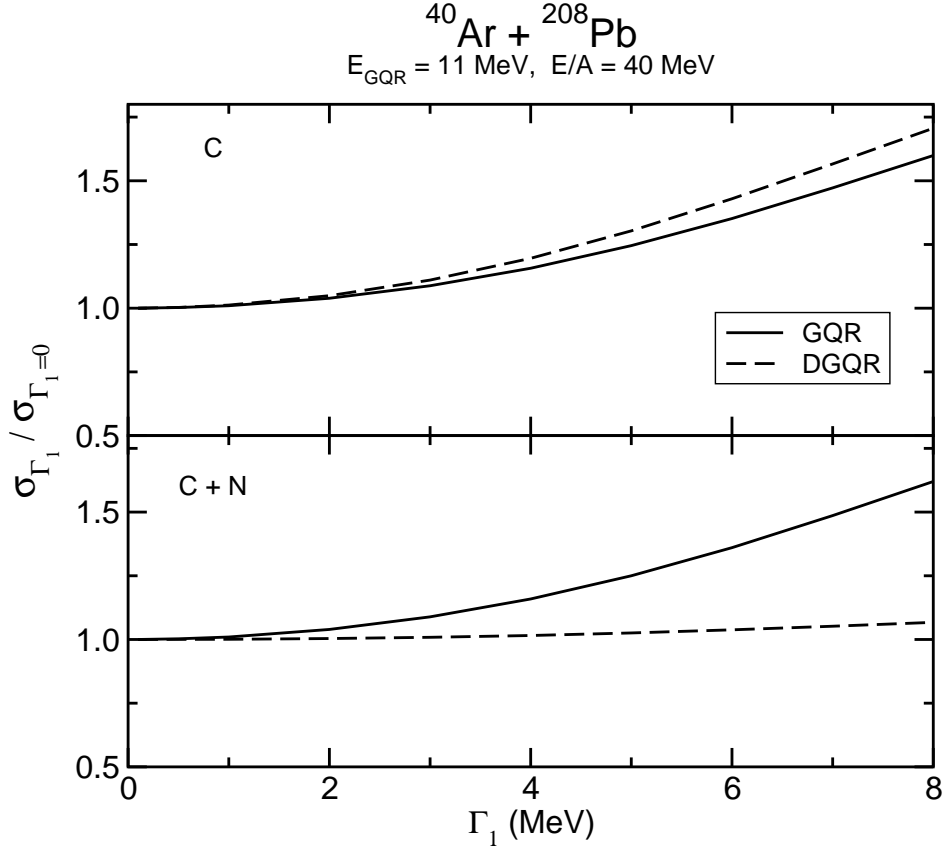}
\end{center}
\caption{Total cross sections for the excitation of GQR (solid line)
and of DGQR (dashed line) as a function of the strength distribution
width $\Gamma_1$ of the single GQR for the reaction $^{40}$Ar +
$^{208}$Pb at $30 MeV/A$. In the graphs we report the cross sections
due to the Coulomb field (upper) and total (lower), each of them
divided by their corresponding value for sharp distribution
($\Gamma_1=0$). We have not reported the nuclear contribution because
the cross sections for both GQR and DGQR do not change appreciably when a finite
distribution is assumed.}
\label{XG}
\end{figure}

\
\begin{figure}[!t]
\begin{center}
\epsfig{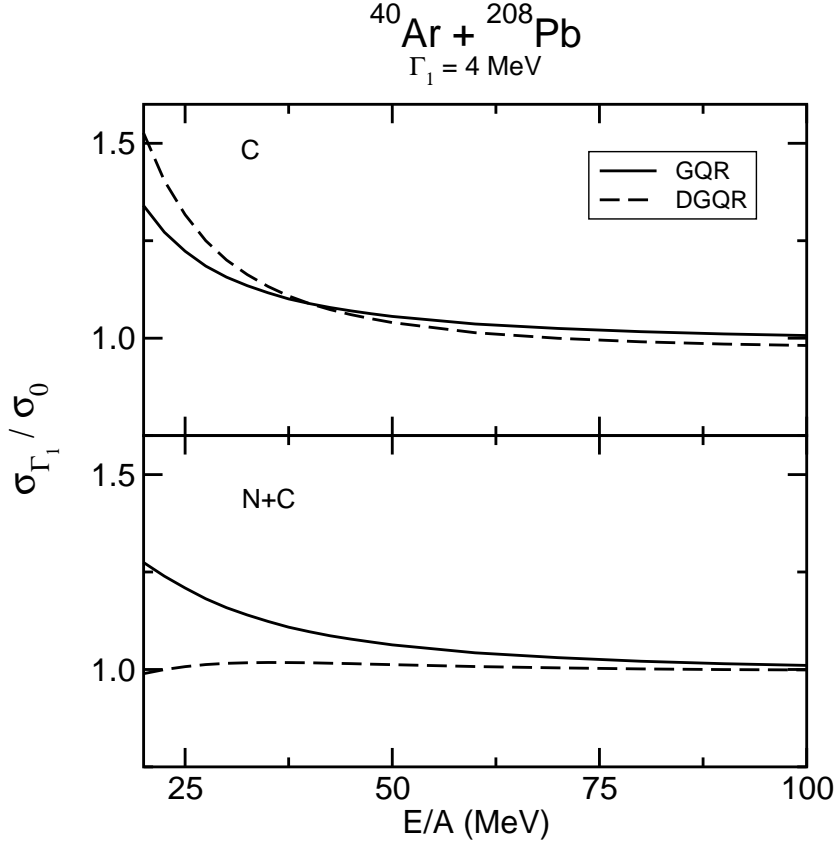}
\end{center}
\caption{Same as fig.~\ref{XG} as a function of the bombarding energy.
The strength distribution width has been chosen equal to 4 MeV.}
\label{XEW}
\end{figure}

\par
We use the same reaction as in Fig.~1 to illustrate the effect of the
reaction Q-values on the transition probabilities.  The dependence of
the single-step inelastic excitation to the one-phonon state of energy
$E_{_{GQR}}$ and the sequential process feeding the double-phonon
state at twice this value are shown in Fig.~2.  As before, the three
curves in each frame display the separate contributions of the Coulomb
and nuclear fields and the combined total.  Two values of the impact
parameter have been chosen specifically to
cover a situation of nuclear ($b$=12 fm) and Coulomb ($b$=13 fm)
dominance. The results show --~even in a linear scale~-- a somewhat
moderate dependence with the frequency of the mode.  This is due to
the relatively high bombarding energy chosen in this example, for
which the time-dependence of both the Coulomb and nuclear excitation
fields are quite well-tuned to the intrinsic response.
\par
There is a qualitative difference in the effective collision time for
Coulomb and nuclear inelastic processes that is worth mentioning.  We
refer to the dependence, for a given bombarding energy, of the
excitation probabilities for the one- and two-phonon states on the
impact parameter. Because of the long range of the formfactors the
change of the effective collision time $\tau$ for Coulomb excitation
follows a different law than the one corresponding to the nuclear
inelastic processes. It can be estimated that $\tau_{_C}/\tau_{_N}
\approx A_\lambda\sqrt{b}$, where the proportionality factor
$A_\lambda$ is a monotonically decreasing function of the
multipolarity $\lambda$.  For all multipolarities, however,
$\tau_{_C}$ is larger than $\tau_{_N}$.  It follows from these
arguments that the adiabatic cut-off function that affects the
transition amplitudes for Coulomb excitation varies significantly over
the large range of impact parameters that contributes to this process.
For the nuclear field a favorable matching between effective collision
times and the intrinsic period of the mode applies, on the other hand,
to most of the relevant partial waves.  This can be understood by
examining Fig.~3, where the probability for excitation of the
one-phonon level is plotted as a function of the energy of the mode
for three impact parameters, $b$=10, 15 and 20 fm.  Two sets of curves
are shown, corresponding to Coulomb and nuclear excitation only.  In
both instances the probabilities are normalized to their values for
$E_{_{GQR}}$=0 MeV to emphasize the different character of the
response.  Notice, for instance, that for $E_{_{GQR}}$=20 MeV the role
of the Coulomb field for $b$=20 fm would be effectively quenched by
two orders of magnitude in spite of its long range. (This is of course
{\rm in addition} to the gradual reduction of the transition
amplitudes caused by the slow $r^{-(\lambda+1)}$ dependence of the
couplings.)
\par
We use Fig.~4 to illustrate the dependence of the excitation
probabilities upon the bombarding energy.  For this we take a value of
$E_{_{GQR}}$=11 MeV, close to the actual excitation energy of the
Giant Quadrupole Resonance in lead.  The impact parameter $b$ is set
to 12.5 fm, which provides the condition in which the importance of
the Coulomb and nuclear excitations become comparable.  Of course it
is also the choice of impact parameter that yields the maximum
(negative) interference between the two reaction mechanisms. What we see
is a rapid increase of the probabilities for the one-phonon and
two-phonon levels up to a bombarding energy of about 50 MeV/nucleon.
After that a gradual decline sets in up to about 400 MeV/nucleon, an
energy beyond which a relativistic formalism must be implemented.  The
trend, however, is not to be significantly altered and, in view of
these results one cannot but wonder about the actual need of
exploiting relativistic bombarding energies to probe the excitation of
double-phonon giant resonances in nuclei.  In principle, and entirely
from an adiabatic point of view, the higher the bombarding energy the
better.  Yet, optimal matching conditions reach saturation and one
cannot ignore the fact that, beyond this point, one can no longer
expect a further enhancement of the excitation probabilities. Quite on
the contrary, the interaction time is effectively reduced up to a
point where (as the figure shows) the excitation of the modes becomes
less and less favored (see caption to Fig.~5 and, also,
ref.~\cite{ber96}).
\par
In Fig.~6 we display the cross section for the excitation of the GQR
and DGQR as a function of the bombarding energy in MeV per nucleon.
This observable quantity combines the effect of all impact parameters
and the plot puts in evidence a quite interesting feature.  Notice
that at all bombarding energies the population of the one-phonon state
is dominated by the Coulomb formfactors.  At the two-phonon level, on
the other hand, it is mostly the nuclear coupling that determines the
outcome.  To understand the origin of this exchange of roles it may be
helpful to re-examine Fig.~1.  We have here to pay attention to the
dependence of the ratio between the probabilities for nuclear and
Coulomb excitation in the relevant range of impact parameters, 11-13
fm.  (To this end the display factor of 30 introduced for the case of
the two-phonon state is of no consequence.)  The enhanced logarithmic
slopes for the DGQR resulting from the squaring of the one-phonon
probabilities suffice to give the leading edge to the nuclear
couplings.  This realization has major consequences insofar as the
global properties of the excitation of the GQR and DGQR is concerned.
In fact, the transition probabilities will inevitably reflect the
different characteristics of the reaction mechanism that it is mostly
responsible for the population of one state or the other.
\par
Q-value considerations have such a pronounced effect on the excitation
probabilities that it is clear that they will play an important role
when one takes into account the sizable width of the one- and
two-phonon states.  Suppose that instead of having the total strength
of the mode at a fixed value $E_{_{GQR}}$ we distribute it with the
profile of a Gaussian distribution of width $\Gamma$.  If the energy
of the mode is quite off the optimal Q-value window one should expect
that the distribution of measured cross sections will follow a quite
different law. In fact, whenever the dynamic response in the vicinity
of $E_{_{GQR}}$ is a rapidly changing function of the energy (see, for
instance, Fig.5 for $E$=10 MeV/A) the experimental distribution will
be significantly distorted and shifted towards lower energies.  We
illustrate this aspect in Fig. 7, where the distribution of Coulomb
excitation probabilities for the one-phonon state, $dP^{(1)}/dE$, is
shown for different impact parameters and bombarding energies.  In
each frame the shaded curve shows the Gaussian distribution of
strength that is the input to the calculation.  Notice that all
distributions have been normalized in order to emphasize the effect of
interest and to eliminate the over-all dependence on $b$ and $E$
discussed earlier.  As it follows from our considerations one can
easily see that the smaller distortion corresponds indeed to the
smaller impact parameters and/or the larger bombarding energies.
\par
The distortion of the line profile at the one-phonon level increases
as a function of the width $\Gamma$, as it is clearly seen in Fig. 8,
where reaction cross sections (i.e. the result of an integration over
impact parameters) are shown for a typical value of the bombarding
energy.  For the larger width $\Gamma$=6 MeV the apparent shift of the
distribution is large enough as to place most of the cross section
outside of the initial range set by the Gaussian curve.  The effect
seems to be more noticeable at the two-phonon level, as shown on the
right-hand-side of the figure.  According to our previous discussion,
it is the Coulomb excitation mechanism that contributes most to the
difference between the strength and cross section profiles.
\par
From the energy distributions displayed in Fig.~8 one can calculate
the total one- and two-phonon cross sections, by integrating over the
excitation energy.  The global effect of the finite width is shown in
Fig.~9, where the total cross sections for different values of the
width are compared with the corresponding values for sharp resonances.
The enhanced excitation in the lower part of the distribution leads to
a global enhancement in the case of the Coulomb field.  As a
consequence, a corresponding enhancement is present in the combined
Coulomb+nuclear case in the one-phonon excitation, which is dominated
by the Coulomb interaction.  On the contrary, being the two-phonon
cross section predominantly due to the nuclear process, no 
appreciable variation
is predicted for this case with finite values of the width.
\par
Since the effect of the width arises from the Q-value kinematic
matching conditions, variations are expected with the bombarding
energy.  In particular one expects that the effects will tend to
vanish at high bombarding energies.  This is illustrated in Fig.~10,
where the total one- and two-phonon cross sections for $\Gamma$ = 4
MeV are compared to the corresponding values for $\Gamma$ = 0 as a
function of the incident energy.
   
\section{Conclusions and remarks}

We have implemented a simple scheme to calculate the excitation probabilities
for the single and double Giant Resonance as a function of several global
parameters such as excitation energies, bombarding energies, width
etc. We have assumed that the colliding nuclei have no structure
except for the presence, in the target, of one and two-phonon
states.  The excitation processes have been calculated within a
semiclassical model and according to perturbation theory. Since both
nuclear and Coulomb interaction are taken into account the cross
sections are calculated by integrating over all range of impact
parameter with an imaginary potential that takes care of the inner
trajectories. The formalism has been applied to the excitation of 
giant resonances in a typical heavy ion reaction, $^{40}$Ar + $^{208}$Pb. 
In our examples, we have limited our calculation to the giant quadrupole resonance.

The role of the nuclear interaction and its interplay with the
long-ranged Coulomb field has been studied. The presence of nuclear
coupling modifies the mechanism excitation of both the GR and the DGR,
the effect being strongly evident in the latter. This has been
ascribed to the difference in the effective collision time which,
together with the qualitative $r$ dependence of the form factors,
produces a different dependence of the transition probabilities on
the reaction Q-value. Hence, the excitation of GR is dominated by the
Coulomb interaction while it is mostly the nuclear coupling which
determines the population of the DGR. 

We have also studied the consequences of the spreading of the strength
distribution of the single giant resonance on the inelastic excitation
of the GR and DGR. Q-value considerations play an important role when
the width of the one- and two-phonon states are considered. Cross
section dependence on both the width of the distribution and the
incident energy has been considered. When compared with the
corresponding values for sharp resonances, the cross sections for GR
and DGR calculated with only the Coulomb field increase as
$\Gamma$ increases. These results are qualitatively similar to the one
obtained in ref.~\cite{wei} where the relativistic Coulomb excitation
of dipole giant resonance (GDR) and double GDR are calculated within a
random matrix theory including the Brink-Axel hypothesis. When the
nuclear interaction is switched on, the enhancement for the single GR
is maintained while the two-phonon cross section presents no variation
with the case of finite value of the width. Also for the dependence on
the incident energies has been found the same trend. This is due to
the fact that the two-phonon cross section is predominantly governed
by nuclear processes.

\section{Acknowledgment}
This work has been partially supported by the Spanish-Italian
agreement between CICyT and INFN and by the Italian MIUR under
contract PRIM 2001-2003 {\it Fisica teorica del nucleo e dei sistemi a
molti corpi}.


\begin{thebibliography}{00}
\bibitem{harak} M. N. Harakeh and A. van Woude, {\it Giant Resonances}
                   (2001 Oxford University Press, New York).
\bibitem{rev} H. Hemling, Prog. Part. Nucl. Phys. {\bf 33} (1994) 729;
                 Ph. Chomaz, and N. Frascaria, Phys. Rep. {\bf 252} (1995) 275;
                 T. Aumann, P. Bortignon, and H. Emling,
                  Annu. Rev. Part. Sci. vol. {\bf 48} (1998) 351.
\bibitem{wam} S. Nishizaki and J. Wambach,Phys. Lett. {\bf B349}
                   (1995) 7;  Phys. Rev {\bf C57} (1998)1515.
\bibitem{pon} V. Yu. Ponomarev, P. F. Bortignon, R. A. Broglia,
                  E. Vigezzi and V. V. Voronov, Nucl. Phys. {\bf A599} (1996) 341c.
\bibitem{ber} C. A. Bertulani and V. Yu. Ponomarev, Phys. Rep. {\bf 321}
                  (1999) 139.
\bibitem{cat}  F. Catara, Ph. Chomaz and N. Van Giai, Phys. Lett {\bf
                 B233} (1989) 6.
\bibitem{pon2} V. Yu. Ponomarev, P. F. Bortignon, R. A. Broglia,
                  and V. V. Voronov, Phys. Rev. Lett. {\bf 85} (2000) 1400.
\bibitem{lan} E. G. Lanza, M. V. Andr\'es, F. Catara, Ph. Chomaz and
                 C. Volpe, Nucl. Phys. {\bf A613} (1996) 445; Nucl. Phys. {\bf A654}
                  (1999) 792c.
\bibitem{vol}  C. Volpe, F. Catara, Ph. Chomaz, M. V.  Andr\'es and
                  E. G. Lanza, Nucl. Phys. {\bf A589} (1995) 521;
                  Nucl. Phys.  {\bf A599} (1996) 347c.
\bibitem{bor} P. F. Bortignon and C. H. Dasso, Phys. Rev. {\bf C56}
                  (1997) 574.
\bibitem{brax} D. Brink, PhD Thesis, Oxford University, 1955,
                  unpublished; P. Axel, Phys. Rev. {\bf 126} (1962)
                  671.
\bibitem{car} B. V. Carlson et al., Ann. Phys. (NY) {\bf 276} (1999)
                 111; Phys. Rev. {\bf C60} (1999) 014604.
\bibitem{wei} J. Z. Gu and H. A. Weidenm\"uller, Nucl. Phys. {\bf
                  A690} (2001) 382.
\bibitem{cat87} F. Catara, Ph. Chomaz and A. Vitturi, Nucl. Phys. {\bf
                      A471} (1987) 661. 
\bibitem{lan2} E. G. Lanza, M. V. Andr\'es, F. Catara, Ph. Chomaz and
                 C. Volpe, Nucl. Phys. {\bf A636} (1998) 452.
\bibitem{and} M. V. Andr\'es, F. Catara, E. G. Lanza, Ph. Chomaz,
                 M. Fallot anf J. A. Scarpaci, Phys. Rev {\bf C65} (2001) 014608.
\bibitem{can} L. F. Canto, A. Romanelli, M. S. Hussein and A. F. R. de
                 Toledo Piza, Phys. Rev. Lett. {\bf 72} (1994) 2147.
\bibitem{ber96} C. A. Bertulani, L. F. Canto, M. S. Hussein and A. F. R. de
                 Toledo Piza, Phys. Rev. {\bf C53} (1996) 334.
\bibitem{ald} K. Alder and  A. Winther, {\it Electromagnetic Excitation}  
                  (North-Holland, Amsterdam, 1975).
\bibitem{win} R. A. Broglia and A. Winther, {\it Heavy Ion Reactions} Vol.I (1981 The
                  Benjamin/Cummings Publishing Company).
\bibitem{land} S. Landowne and A. Vitturi, in {\it Treatise on Heavy Ion
                 Science},  Ed. D. A. Bromley, vol. 1, p.355.

\end{thebibliography}
\end{document}